\documentclass[prd,preprint,nofootinbib]{revtex4-1} 
\usepackage{latexsym,verbatim}
\usepackage{amssymb}
\usepackage{amsfonts}
\usepackage{amsmath,color}
\usepackage[draft]{graphicx}
\usepackage{bm}
\usepackage[utf8]{inputenc}
\usepackage[T1]{fontenc}

\usepackage{hyperref}

\hypersetup{
    colorlinks=true,
    linkcolor=red,
    citecolor=green,
    filecolor=magenta,      
    urlcolor=cyan,
    pdftitle={GWs and Fermi},
    pdfpagemode=FullScreen,}

\def\mb#1{\mathbf{#1}}

\def\ber{\begin{eqnarray}}
\def\eer{\end{eqnarray}}
\def\beq{\begin{equation}}
\def\eeq{\end{equation}}

\def\rmd{{\rm d}}

\def\ed{\end{document}}

\def\sT{\sin \left(\omega T \right)}
\def\cT{\cos \left(\omega T \right)}

\newcommand{\ppar}[2]{\frac{\partial #1}{\partial #2}}

\def\fp{f^{+}}
\def\fc{f^{\times}}

\makeatletter
\let\jnl@style=\rm
\def\ref@jnl#1{{\jnl@style#1}}

\def\aj{\ref@jnl{AJ}}                   % Astronomical Journal
\def\actaa{\ref@jnl{Acta Astron.}}      % Acta Astronomica
\def\araa{\ref@jnl{ARA\&A}}             % Annual Review of Astron and Astrophys
\def\apj{\ref@jnl{ApJ}}                 % Astrophysical Journal
\def\apjl{\ref@jnl{ApJ}}                % Astrophysical Journal, Letters
\def\apjs{\ref@jnl{ApJS}}               % Astrophysical Journal, Supplement
\def\ao{\ref@jnl{Appl.~Opt.}}           % Applied Optics
\def\apss{\ref@jnl{Ap\&SS}}             % Astrophysics and Space Science
\def\aap{\ref@jnl{A\&A}}                % Astronomy and Astrophysics
\def\aapr{\ref@jnl{A\&A~Rev.}}          % Astronomy and Astrophysics Reviews
\def\aaps{\ref@jnl{A\&AS}}              % Astronomy and Astrophysics, Supplement
\def\azh{\ref@jnl{AZh}}                 % Astronomicheskii Zhurnal
\def\baas{\ref@jnl{BAAS}}               % Bulletin of the AAS
\def\bac{\ref@jnl{Bull. astr. Inst. Czechosl.}}
\def\caa{\ref@jnl{Chinese Astron. Astrophys.}}
\def\cjaa{\ref@jnl{Chinese J. Astron. Astrophys.}}
\def\icarus{\ref@jnl{Icarus}}           % Icarus
\def\jcap{\ref@jnl{J. Cosmology Astropart. Phys.}}
\def\jrasc{\ref@jnl{JRASC}}             % Journal of the RAS of Canada
\def\memras{\ref@jnl{MmRAS}}            % Memoirs of the RAS
\def\mnras{\ref@jnl{MNRAS}}             % Monthly Notices of the RAS
\def\na{\ref@jnl{New A}}                % New Astronomy
\def\nar{\ref@jnl{New A Rev.}}          % New Astronomy Review
\def\pra{\ref@jnl{Phys.~Rev.~A}}        % Physical Review A: General Physics
\def\prb{\ref@jnl{Phys.~Rev.~B}}        % Physical Review B: Solid State
\def\prc{\ref@jnl{Phys.~Rev.~C}}        % Physical Review C
\def\prd{\ref@jnl{Phys.~Rev.~D}}        % Physical Review D
\def\pre{\ref@jnl{Phys.~Rev.~E}}        % Physical Review E
\def\prl{\ref@jnl{Phys.~Rev.~Lett.}}    % Physical Review Letters
\def\pasa{\ref@jnl{PASA}}               % Publications of the Astron. Soc. of Australia
\def\pasp{\ref@jnl{PASP}}               % Publications of the ASP
\def\pasj{\ref@jnl{PASJ}}               % Publications of the ASJ
\def\rmxaa{\ref@jnl{Rev. Mexicana Astron. Astrofis.}}%
\def\qjras{\ref@jnl{QJRAS}}             % Quarterly Journal of the RAS
\def\skytel{\ref@jnl{S\&T}}             % Sky and Telescope
\def\solphys{\ref@jnl{Sol.~Phys.}}      % Solar Physics
\def\sovast{\ref@jnl{Soviet~Ast.}}      % Soviet Astronomy
\def\ssr{\ref@jnl{Space~Sci.~Rev.}}     % Space Science Reviews
\def\zap{\ref@jnl{ZAp}}                 % Zeitschrift fuer Astrophysik
\def\nat{\ref@jnl{Nature}}              % Nature
\def\iaucirc{\ref@jnl{IAU~Circ.}}       % IAU Cirulars
\def\aplett{\ref@jnl{Astrophys.~Lett.}} % Astrophysics Letters
\def\apspr{\ref@jnl{Astrophys.~Space~Phys.~Res.}}
\def\bain{\ref@jnl{Bull.~Astron.~Inst.~Netherlands}}
\def\fcp{\ref@jnl{Fund.~Cosmic~Phys.}}  % Fundamental Cosmic Physics
\def\gca{\ref@jnl{Geochim.~Cosmochim.~Acta}}   % Geochimica Cosmochimica Acta
\def\grl{\ref@jnl{Geophys.~Res.~Lett.}} % Geophysics Research Letters
\def\jcp{\ref@jnl{J.~Chem.~Phys.}}      % Journal of Chemical Physics
\def\jgr{\ref@jnl{J.~Geophys.~Res.}}    % Journal of Geophysics Research
\def\jqsrt{\ref@jnl{J.~Quant.~Spec.~Radiat.~Transf.}}
\def\memsai{\ref@jnl{Mem.~Soc.~Astron.~Italiana}}
\def\nphysa{\ref@jnl{Nucl.~Phys.~A}}   % Nuclear Physics A
\def\physrep{\ref@jnl{Phys.~Rep.}}   % Physics Reports
\def\physscr{\ref@jnl{Phys.~Scr}}   % Physica Scripta
\def\planss{\ref@jnl{Planet.~Space~Sci.}}   % Planetary Space Science
\def\procspie{\ref@jnl{Proc.~SPIE}}   % Proceedings of the SPIE

\makeatother

\begin{document}

\author{Matteo Luca Ruggiero}
\email{matteoluca.ruggiero@unito.it}
\affiliation{Dipartimento di Matematica ``G.Peano'', Universit\`a degli studi di Torino, Via Carlo Alberto 10, 10123 Torino, Italy}
\affiliation{INFN - LNL , Viale dell'Universit\`a 2, 35020 Legnaro (PD), Italy}
\date{\today}

\title{A note on the description of plane gravitational waves in Fermi coordinates}

\begin{abstract}
We use the formalism of Fermi coordinates to describe the interaction of a plane gravitational wave in the proper detector frame. In doing so, we emphasize that in this frame the action of the gravitational wave can be explained in terms of a gravitoelectromagnetic analogy. In particular, up to linear displacements from the reference world-line, the effects of the wave on test masses can be described in terms of a Lorentz-like force equation. 
In this framework we focus on the effects on time measurements provoked by the passage of the wave, and evaluate their order of magnitude. Eventually, we calculate the expression of the local spacetime metric in cylindrical coordinates adapted to the symmetries of the gravitational field and show its relevance in connection with the helicity-rotation coupling.
\end{abstract}

\maketitle

 %------------------------Section-------------------------
\section{Introduction} \label{sec:intro}
%------------------------Section-------------------------

Since 2015, the year when the first direct detection took place \cite{PhysRevLett.116.061102}, we have been in the era of gravitational waves (GWs) astronomy; as a matter of fact, the recent announcement from the NANOGrav collaboration \cite{NANOGrav:2023gor}, which suggested the evidence for a stochastic gravitational-wave background, is just the latest important success in this research field. GWs are to all effects a new tool to explore the Universe and they allow to harness the potentiality of  multi-messenger astronomy, according to which a given astrophysical source can be detected and studied by means of different messengers \cite{burns2019opportunities}. 

GWs are typically described using the transverse and traceless (TT)  tensor, which allows to introduce the so-called \textit{TT coordinates or TT frame}  (see e.g.\citet{Rakhmanov_2014} for a thorough discussion on the various coordinates used to describe the interaction with GWs).  Actually, these coordinates are used because they emphasize the physical degrees of freedom and do not contain gauge-depending  terms. As a matter of fact, however, they  lack a direct physical meaning since they are not simply related to measurable quantities \cite{flanagan2005basics}: for instance,  the TT coordinates of a test mass in the field of a gravitational wave do not change  (see e.g. \citet{Ruggiero:2021qnu});  of course, things are different if we consider the physical distance between test masses, which is modified by the passage of the wave, and is the base of the interferometric detection.

There is, however, another approach that can be used to describe the interaction between GWs and detectors, and this is the so-called \textit{proper detector frame} \cite{maggiore2007gravitational} which is based on the use of \textit{Fermi coordinates}. The latter  are a quasi-Cartesian coordinates system that can be build in the neighbourhood of the world-line of an observer, and their definition depends both on the background field where the observer is moving and, also, on his  motion. Fermi coordinates are  defined, by construction, as scalar invariants \cite{synge1960relativity}; they have a concrete meaning, since they are the coordinates an observer would naturally use to make space and time measurements in the vicinity of his world-line. 

Using Fermi coordinates we showed that the interaction of a plane gravitational wave can be described in terms of a \textit{gravitoelectromagnetic} analogy \cite{Ruggiero_2020,Ruggiero:2021qnu,Ruggiero:2022gzl}: in other words, the effects on test masses in the detector frame can be seen as the action of a gravitoelectric and a gravitomagnetic field. To this end, it is useful to remember that analogies with electromagnetism naturally arise in General Relativity in different contexts (see e.g. \citet{Ruggiero:2002hz,Mashhoon:2003ax,Costa:2012cw,Ruggiero:2023ker}). This approach allows to emphasize the fact that current detectors reveal the interaction of test masses with the gravitoelectric components of the wave, but there are also gravitomagnetic interactions that could be used to detect the effect of GWs on moving masses  and spinning particles \cite{biniortolan2017,Ruggiero_2020b}.

{Gravitomagnetic effects are also relevant in the emission of GWs: as discussed by \citet{rev2}, the waveforms are affected by the higher order gravitomagnetic corrections to the equation of motion of the orbits of a binary system \cite{rev1}; in particular, an accurate analysis shows that these effects are important to obtain a peculiar characterization of the dynamics of the binary system.}

In this paper we further develop this analogy and exploit it to investigate other effects that are due to the passage of the wave. The interest in developing this formalism comes from some recent papers which focus on the possibility to test GWs physics with experimental setups that are different from interferometers, such as accelerators \cite{acceleratori,acceleratori2}; more generally speaking, the proper detector frame is naturally used to describe possible GWs effects in storage rings \cite{storage} and microcavities  \cite{Berlin:2021txa}.

The paper is organised as follows: in Section \ref{sec:gem} we introduce the gravitoelectromagnetic analogy in Fermi coordinates, while in Section \ref{sec:time} we evaluate the impact of GWs on time measurements; Section \ref{sec:metric} is devoted to the description of the Fermi metric in cylindrical coordinates, while discussions and conclusions are in Section \ref{sec:disconc}.

 %------------------------Section-------------------------
\section{Gravitoelectric and gravitomagnetic fields arising from Fermi coordinates} \label{sec:gem}
%------------------------Section-------------------------

Fermi coordinates enable to describe the expression of the spacetime metric in the vicinity of a given world-line, which in practice can be thought of as the world-line of the laboratory frame, where the detector is placed. In general, this expression depends both on the properties of the local reference frame (the world-line acceleration and the tetrad rotation)  and on the spacetime curvature, through the Riemann curvature tensor (see e.g. \citet{Ni:1978di,Li:1979bz,1982NCimB..71...37F,marzlin}, \citet{Ruggiero_2020}).  For the sake of simplicity, since here we focus on GWs, we neglect the inertial effects in the definition of the metric elements: in other words we  consider a geodesic and non rotating frame. Accordingly, using  Fermi coordinates $(cT,X,Y,Z)$, up to quadratic displacements $|X^{i}|$ from the reference world-line, the line element turns out to be\footnote{Latin indices refer to space coordinates, while Greek indices to spacetime ones. We will use bold-face symbols like  $\mb W$ to refer to vectors in the Fermi frame; the spacetime signature is $+2$ in our convention.} (see e.g.  \citet{manasse1963fermi,MTW})
\beq
ds^{2}=-\left(1+R_{0i0j}X^iX^j \right)c^{2}dT^{2}-\frac 4 3 R_{0jik}X^jX^k cdT dX^{i}+\left(\delta_{ij}-\frac{1}{3}R_{ikjl}X^kX^l \right)dX^{i}dX^{j}. \label{eq:mmmetric}
\eeq
Here  $R_{\alpha \beta \gamma \delta}(T)$ is the projection of the 
Riemann curvature tensor on the orthonormal tetrad $e^{\mu}_{(\alpha)}(\tau)$ of the
reference observer, parameterized by the proper time\footnote{In $e^{\mu}_{(\alpha)}$ tetrad indices like $(\alpha)$ are within parentheses, while  $\mu$ is a  background spacetime index; however, for the sake of simplicity, we drop here and henceforth parentheses to refer to tetrad indices, which are the only ones used.} $\tau$: $\displaystyle R_{\alpha \beta \gamma \delta}(T) = R_{\alpha \beta \gamma \delta}(\tau)=R_{\mu\nu \rho
\sigma}e^\mu_{(\alpha)}(\tau)e^\nu_{(\beta)}(\tau)e^\rho_{(\gamma)}(\tau)e^\sigma 
_{(\delta)}(\tau)$ and it is evaluated along the reference geodesic, where $T=\tau$ and $ X^{i}=0$.

By setting
\[
\frac{\Phi}{c^{2}}=\frac{g_{00}+1}{2} \quad \frac{\Psi_{ij}}{c^{2}}=\frac{g_{ij}-\delta_{ij}}{2} \quad \frac{A_{i}}{c^{2}}=-\frac{g_{0i}}{2},
\]
the above metric can be written in the form
\beq
\mathrm{d} s^2= -c^2 \left(1-2\frac{\Phi}{c^2}\right)\rmd T^2 -\frac4c A_{i}\rmd X^{i}\rmd T  +
 \left(\delta_{ij}+2\frac{\Psi_{ij}}{c^2}\right)\rmd X^i \rmd X^j\ , \label{eq:weakfieldmetric11}
\eeq
with the following definitions
\begin{eqnarray}
\Phi (T, { X^{i}})&=&-\frac{c^{2}}{2}R_{0i0j}(T )X^iX^j, \label{eq:defPhiG}\\
A^{}_{i}(T ,{X^{i}})&=&\frac{c^{2}}{3}R_{0jik}(T )X^jX^k, \label{eq:defAG}\\
\Psi_{ij} (T, {X^{i}}) & = & -\frac{c^{2}}{6}R_{ikjl}(T)X^{k}X^{l}, \label{eq:defPsiG}
\end{eqnarray}
where $\Phi$ and $A_{i}$ are, respectively, the \textit{gravitoelectric} and \textit{gravitomagnetic} potential, and $\Psi_{ij}$ is the perturbation of the spatial metric \cite{Ruggiero_2020,Ruggiero:2021uag}. Notice that the line element (\ref{eq:weakfieldmetric11}) is a perturbation of flat Minkowski spacetime; in other words $|\frac{\Phi}{c^{2}}| \ll 1$, $|\frac{\Psi_{ij}}{c^{2}}| \ll 1$, $|\frac{A_{i}}{c^{2}}| \ll 1$.

We want to apply this formalism to the field of a plane gravitational wave. Before doing that, we briefly recall the standard approach to the description of plane gravitational waves using the TT coordinates. 
We start from Einstein's equations 
\beq
G_{\mu\nu}=\frac{8\pi G}{c^{4}}T_{\mu\nu}, \label{eq:einstein0}
\eeq
and we suppose that the spacetime metric $g_{\mu\nu}$ is in the form $\displaystyle g_{\mu\nu}=\eta_{\mu\nu}+h_{\mu\nu}$, where $|h_{\mu\nu}|\ll 1$ is a small perturbation of the Minkowski tensor $\eta_{\mu\nu}$ of flat spacetime. Setting $\bar h_{\mu\nu}=h_{\mu\nu}-\frac 1 2 \eta_{\mu\nu}h$, with $h=h^{\mu}_{\mu}$, Einstein's field equations (\ref{eq:einstein0}) in the Lorentz gauge $\displaystyle \partial_{\mu} \bar h^{\mu\nu}=0$ (where $\displaystyle \partial_{\mu} = \frac{\partial}{\partial x^{\mu}}$) turn out to be
\beq
\square \bar h_{\mu\nu}=-\frac{16\pi G}{c^{4}}T_{\mu\nu}, \label{eq:einstein2}
\eeq
where $\square = \partial_{\mu}\partial^{\mu}=\nabla^{2}-\frac{1}{c^{2}}\frac{\partial}{\partial t^{2}}$. The vacuum (i.e. $T_{\mu\nu}=0$) solutions of Eq. (\ref{eq:einstein2}) are GWs propagating in empty space. Using TT coordinates $(ct_{\mathrm{TT}},x_{\mathrm{TT}},y_{\mathrm{TT}},z_{\mathrm{TT}}$) \cite{{Rakhmanov_2014}} the solutions for a wave propagating along the $x$ direction are given by 
\beq
\bar h_{\mu\nu}=-\left(h^{+}e^{+}_{\mu\nu}+h^{\times}e^{\times}_{\mu\nu} \right), \label{eq:gwsol1}
\eeq
{with
\beq
h^{+}=A^{+}\cos \left(\omega t_{\mathrm{TT}}-kx_{\mathrm{TT}} +\phi^{+} \right), \quad h^{\times}=A^{\times}\cos \left(\omega t_{\mathrm{TT}}-kx_{\mathrm{TT}} +\phi^{\times}\right), \label{eq:gwsol20}
\eeq
where $\phi^{+}, \phi^{\times}$ are constants, and
\beq
e^{+}_{\mu\nu}=\left[\begin{array}{cccc}0 & 0 & 0 & 0 \\0 & 0 & 0 & 0 \\0 & 0 & 1 & 0 \\0 & 0 & 0 & -1\end{array}\right], \quad e^{\times}_{\mu\nu}=\left[\begin{array}{cccc}0 & 0 & 0 & 0 \\0 & 0 & 0 & 0 \\0 & 0 & 0 & 1 \\0 & 0 & 1 & 0\end{array}\right] \label{eq:gwsol3}
\eeq 
are the linear polarization tensors of the wave. In the above definitions, $A^{+}, A^{\times}$ are the amplitude of the wave in the two polarization states, $\phi^{+}, \phi^{\times}$ the corresponding phases,  while $\omega$ is the frequency and $k$ the wave number, so that the wave four-vector is $\displaystyle k^{\mu}=\left(\frac \omega c, k, 0, 0 \right)$, with $k^{\mu}k_{\mu}=0$.  The two linear polarizations states can be added with phase difference of $\pm \pi/2$ to get circularly polarized waves. We will use 
\beq
h^{+}=A^{+}\sin \left(\omega t_{\mathrm{TT}}-kx_{\mathrm{TT}}  \right), \quad h^{\times}=A^{\times}\cos \left(\omega t_{\mathrm{TT}}-kx_{\mathrm{TT}}\right), \label{eq:gwsol2}
\eeq
thus fixing the phase difference: accordingly, circular polarization corresponds to the condition $A^{+}=\pm A_{\times}$. In conclusion, in TT coordinates the spacetime element is given by
\beq
ds^2= -c^{2}dt_{\mathrm{TT}}^2+dx_{\mathrm{TT}}^2 +(1-h^{+})dy_{\mathrm{TT}}^2 +(1+h^{+})dz_{\mathrm{TT}}^2 -2h_{\times} dy_{\mathrm{TT}} dz_{\mathrm{TT}}\,. \label{eq:TTmetrica}
\eeq

We remember that up to linear order in the perturbation $h_{\mu\nu}$, we can write the following expressions for the Riemann tensor \cite{MTW}:
\beq
R_{ikjl}=\frac 1 2 \left(h_{il,jk}+h_{kj,li}-h_{kl,ji}-h_{ij,lk} \right) \label{eq:riemann0}
\eeq
and
\beq
R_{ij0l}=\frac 1 2 \left( h_{il,j0}-h_{jl,i0} \right). \label{eq:riemann1}
\eeq
We can exploit the gauge invariance in linear approximation \cite{straumann2013applications}  and use the above expressions to calculate the Riemann tensor in Fermi coordinates; notice that in the metric (\ref{eq:mmmetric}) the Riemann tensor is evaluated along the reference world-line: accordingly, after calculating the components of Riemann tensor using Eq. (\ref{eq:gwsol2}), we set $X^{i}=0$. We point out that the expression of the metric tensor is obtained in the large wavelength limit, which means that the typical dimension $L$ of the frame  is negligible with respect to the wavelength $\lambda$; more accurate expressions can be obtained, which contains higher order terms in the small parameter $\displaystyle \epsilon = \frac{L}{\lambda}$  (see e.g. \citet{Ruggiero:2022gzl} and references therein). Actually, the series expansion can be exactly summed to obtain a compact form \cite{1982NCimB..71...37F,Berlin:2021txa}.

If we define the functions
\beq
f^{+}=\frac 1 2 A^{+}\sT, \quad f^{\times}=\frac 1 2 A^{\times} \cT \label{def:fpfc}
\eeq
the gravitoelectric potential (\ref{eq:defPhiG}) is written as
\beq
\Phi=\frac{\omega^{2}}{2}\left[ f^{+} Y^{2}+2f^{\times}\ YZ-f^{+} Z^{2} \right], \label{eq:defPhicomp}
\eeq

%OLD
%\beq
%\Phi=\frac{\omega^{2}}{4}\left[ A^{+}\sT Y^{2}+2A^{\times}\cT YZ-A^{+}\sT Z^{2} \right], \label{eq:defPhicomp}
%\eeq

while the components of the gravitomagnetic potential (\ref{eq:defAG})  are

\begin{eqnarray}
A_{X}&=&\frac{\omega^{2}}{3} \left[f^{+} (Y^{2}-Z^{2})+2f^{\times}\ ZY \right], \label{eq:defAX} \\
A_{Y}&=& \frac{\omega^{2}}{3} \left[-f^{+} YX-f^{\times} ZX \right], \label{eq:defAY} \\
A_{Z}&=& \frac{\omega^{2}}{3} \left[-f^{\times} YX+f^{+} XZ \right].  \label{eq:defAZ}
\end{eqnarray}

%OLD
%\begin{eqnarray}
%A_{X}&=&\frac{\omega^{2}}{6} \left[A^{+}\sT (Y^{2}-Z^{2})+2A^{\times}\cT ZY \right], \label{eq:defAX} \\
%A_{Y}&=& \frac{\omega^{2}}{6} \left[-A^{+}\sT YX-A^{\times}\cT ZX \right], \label{eq:defAY} \\
%A_{Z}&=& \frac{\omega^{2}}{6} \left[-A^{\times}\cT YX+A^{+}\sT XZ \right].  \label{eq:defAZ}
%\end{eqnarray}

In addition, starting from the definition
\beq
\Psi_{ij} (T, {X^{i}})  =  -\frac{c^{2}}{6}R_{ikjl}(T)X^{k}X^{l}, \label{eq:defPsiGbis}
\eeq
we explicitly calculate the metric components:
\begin{eqnarray}
\Psi_{XX} & = & -\frac 1 6 \omega^{2} \left(-\fp Y^{2}+\fp Z^{2}-2\fc YZ \right), \nonumber \\ 
\Psi_{XY} & = & -\frac 1 6 \omega^{2} \left(\fp YX+\fp ZX \right), \nonumber \\
\Psi_{XZ} & = & -\frac 1 6 \omega^{2} \left(\fc YX-\fp ZX \right), \nonumber \\
\Psi_{YY} & = & -\frac 1 6 \omega^{2} \left(-\fp X^{2} \right), \nonumber \\
 \Psi_{YZ} & = & -\frac 1 6 \omega^{2} \left(-\fc X^{2} \right), \nonumber \\
 \Psi_{ZZ} & = & -\frac 1 6 \omega^{2} \left(\fp X^{2} \right). \label{eq:defPsi}
\end{eqnarray}

If, from the above expression, we define the {gravitomagnetic field}
\beq
\mb B= \bm \nabla \wedge \mb A, \label{eq:defB}
\eeq
and the {gravitoelectric field}
\beq
 \quad \mb E= -\bm \nabla \Phi-\frac{2}{c} \ppar{\mb A}{T}, \label{eq:defEtime1}
\eeq
the geodesic equations can be written in the form \cite{Ruggiero:2023ker}
\beq
\frac{\rmd^{2} X^i}{\rmd T^{2}}=-{ E}^{i}-2 \left(\frac{{\mathbf V}}{c}\times {\mathbf B}\right)^{i}-2\frac{V^{j}}{c} \frac{\partial \Psi_{ij} }{c\partial T}-\frac{V^{i}}{c} \frac{\partial \Phi }{c\partial T}.  \label{eq:lor2}
\eeq
In particular,  this equation defines the motion of a test mass with respect to the reference observer. Consequently, all quantities involved are \textit{relative} to the reference observer at the origin of the frame. We see that this equations is not in a Lorentz-like form if the fields are not static, due to the presence of the last terms which contain time-derivatives \cite{Ruggiero:2021uag}. However, since both terms - according to the definitions (\ref{eq:defPhiG}) and (\ref{eq:defPsiG}) - are quadratic in the displacements from the reference world-line, if we confine ourselves to linear order we obtain the Lorentz-like force
\beq
\frac{\rmd^{2} \mb X}{\rmd T^{2}}=-{ \mb E}-2 \left(\frac{{\mathbf V}}{c}\times {\mathbf B}\right);  \label{eq:lor2lin}
\eeq
which can be used to describe the action of the time-varying field of a gravitational wave; notice that in this case $\mb E=-\bm \nabla \Phi$.

 %------------------------Section-------------------------
\section{Differential time measurements in the field of plane GWs} \label{sec:time}
%------------------------Section-------------------------

Using the above definitions for the gravitoelectric  (\ref{eq:defPhicomp}) and gravitomagnetic potentials (\ref{eq:defAX})-(\ref{eq:defAZ}) and spatial curvature (\ref{eq:defPsi}) we can explicitly write the metric (\ref{eq:weakfieldmetric11}).

We remember that $T$ represents the proper time in the Fermi frame, i.e. the time measured by a clock at rest on the reference world-line; because of its tidal nature, the GWs field produces differential effects for observers at different locations.  This fact is reflected on how different clocks behave with respect to the reference clock: in other words, time measurements are modified by the passage of GWs. 

To begin with, we consider the effects on clocks at rest at different locations. The relation between the infinitesimal proper time difference $dT$ measured by the reference clock, and the corresponding one $d\tau$ measured by another clock at rest in the field of the gravitational wave is given by
\beq
d\tau=\sqrt{\left(1-2\frac{\Phi}{c^2}\right)}  \rmd T. \label{eq:dTfisso1}
\eeq

On integrating from $T=0$ to $T=\Delta T$, we obtain the corresponding increment of the proper time $\Delta \tau$:
\beq
\Delta \tau=\Delta T+\frac{(Z^{2}-Y^{2})A^{+}\omega}{4c^{2}}\left(1-\cos \omega \Delta T \right)-\frac{YZ A^{\times}\omega}{2c^{2}}\sin \omega \Delta T. \label{eq:dTfisso2}
\eeq

We remember that the metric (\ref{eq:weakfieldmetric11}) describes the deformation of  flat spacetime due to the passage of GWs; in other words, spacetime is flat before and after the passage of the GWs. In particular, in flat spacetime clocks are naturally synchronized by Einstein's procedure. However, the synchronization of clocks in a metric in the form (\ref{eq:weakfieldmetric11}) requires  attention, and it is well known \cite{landau2013classical} that a desynchronization effect arises: in other words the passage of the wave produces a desynchronization effect that can be written as 
\beq
c\Delta T_{synch}=-\frac{g_{0i}dX^{i}}{g_{00}} \simeq - g_{0i}dX^{i} =\frac{2}{c^{2}}A_{i}dX^{i}. \label{eq:desynch1}
\eeq
On using the above expressions we obtain, integrating\footnote{Notice that this integration is path depending, since $\bm \nabla \times \mb A = \bm B \neq 0$.} from $0.. X, 0..Y, 0.. Z$, which means that the confrontation is between a clock at the origin and another one at  the generic location $(X,Y,Z)$
\beq
\Delta T_{synch}=\frac{1}{6}\frac{A^{+}\omega^{2} X( Y^{2}- Z^{2})}{c^{3}} \sin \omega T. \label{eq:desynch2}
\eeq
In particular, this effect is null in the plane orthogonal to the propagation direction where $X=0$, and along the propagation direction itself ($Y=Z=0$). 

Now we consider the case of a clock rotating with constant rotation rate $\Omega$, at radius $R$ in the plane $X=0$; we calculate the proper time  $\Delta \tau$ measured along its world line and compare it with the corresponding proper time interval $\Delta T$ measured by the reference clock. 
We set  $\displaystyle \gamma=\sqrt{\frac{1}{1-\frac{\Omega^{2}R^{2}}{c^{2}}}}$. 
Since $Y=R \cos \Omega T, Z=R \sin \Omega T$, we obtain for circularly polarized GWs ($A^{+}=\pm A^{\times}$)
\beq
\Delta \tau=\frac{\Delta T}{\gamma}+\frac{\gamma \omega^{2}R^{2}A^{+}\left[\cos\left(2\Omega\pm\omega\right)\Delta T-1 \right]}{4\left(2\Omega\pm\omega\right)c^{2}}. \label{eq:circ1}
\eeq
In particular, the first term in (\ref{eq:circ1}) is purely kinematic, while the other is determined by the field of the GWs.
We see that the above expression is singular when $\omega=\mp 2\Omega$: in this case we calculate the corresponding expression for general polarization, and we obtain
\beq
\Delta \tau=\frac{\Delta T}{\gamma}-\frac{\gamma \Omega R^{2}\left(A^{\times}\mp A^{+} \right)\sin^{2}\Omega \Delta T \cos^{2} \Omega \Delta T}{c^{2}}. \label{eq:circ2}
\eeq
We notice that the second term in Eq. (\ref{eq:circ2}) is absent for circularly polarized GWs.

 %------------------------Section-------------------------
\section{Metric in cylindrical coordinates} \label{sec:metric}
%------------------------Section-------------------------

In order to further emphasize the features of the field of  GWs, it is useful to write the metric (\ref{eq:weakfieldmetric11}) in cylindrical coordinates, adapted to the spacetime symmetries: to this end, we consider the coordinate transformation   from Cartesian $(cT,X,Y,Z)$ coordinates to cylindrical coordinates $(ct,x,r,\varphi)$ defined by
\begin{eqnarray}
T & = & t, \label{eq:cordtrasT} \\
X & = & x, \label{eq:cordtrasX} \\
Y & = & r \cos \varphi \label{eq:cordtrasY} \\ 
Y & = & r \sin \varphi \label{eq:cordtrasZ} \\ 
\end{eqnarray}
In particular the symmetry axis coincides with the propagation direction of the wave. From this coordinate transformation it is possible to define the Jacobian matrix $\displaystyle \frac{\partial X^{\alpha}}{\partial x^{\mu}}$:
\beq
 \frac{\partial X^{\alpha}}{\partial x^{\mu}} \doteq \left(\begin{array}{ccc}\frac{\partial Y}{\partial r } & \frac{\partial Z}{\partial r } & 1 \\\frac{\partial Y}{\partial \varphi } & \frac{\partial Z}{\partial \varphi } & 0 \\0 & 0 & 1\end{array}\right) \label{eq:jactrasf}
\eeq
where
\begin{eqnarray}
\frac{\partial Y}{\partial r } & = & \cos \varphi, \label{eq:Yr} \\
\frac{\partial Z}{\partial r } & = & \sin \varphi, \label{eq:Zr} \\
\frac{\partial Y}{\partial \varphi } & = & -r \sin \varphi, \label{eq:Yphi} \\
\frac{\partial Y}{\partial \varphi } & = & r \cos \varphi. \label{eq:Zphi}
\end{eqnarray}

Accordingly, the metric $g'_{\alpha\beta}$ in the new coordinates  is defined by
\beq
g'_{\alpha\beta}= \frac{\partial X^{\mu}}{\partial x^{\alpha}} \frac{\partial X^{\nu}}{\partial x^{\beta}} g_{\mu\nu}, \label{eq:trasmetrica1}
\eeq
where the old metric is calculated in terms of the new coordinates  $g_{\mu\nu}=g_{\mu\nu}(X^{\mu}(x^{\alpha}))$. 

In particular, we are interested in the case of circularly polarized GWs, so that $A^{+}=\pm A^{\times}$. Accordingly, we get the following expression of the metric elements in the new coordinates:

\begin{eqnarray}
g'_{00} & = & -c^{2}\left[1-\frac{\omega^{2}r^{2}A^{+}}{2c^{2}}\sin\left(2\varphi\pm\omega t \right) \right], \label{eq:gp00}  \\
g'_{0r} & = & \frac{\omega^{2}xr A^{+}}{3c^{2}}\left[\sin\left(2\varphi\pm\omega t\right)\right], \label{eq:gp0r}\\
g'_{0\varphi} & = & \frac{\omega^{2}xr^{2} A^{+}}{3c^{2}}\left[\cos\left(2\varphi\pm\omega t\right)\right], \label{eq:gp0phi}\\ 
g'_{0x} & = & - \frac{\omega^{2}r^{2} A^{+}}{3c^{2}}\left[\sin\left(2\varphi\pm\omega t\right)\right], \label{eq:gp0x} \\
g'_{rr} & = &1+ \frac{\omega^{2}x^{2} A^{+}}{6c^{2}}\left[\sin\left(2\varphi\pm\omega t\right)\right], \label{eq:gprr} \\
g'_{\varphi\varphi} & = &r^{2}- \frac{\omega^{2}x^{2} r^{2} A^{+}}{6c^{2}}\left[\sin\left(2\varphi\pm\omega t\right)\right], \label{eq:gpphiphi} \\
g'_{r\phi} & = &\frac{\omega^{2} x^{2} r A^{+}}{6c^{2}}\left[\cos\left(2\varphi\pm\omega t\right)\right],\label{eq:gprphi}  \\
g'_{xr} & = &-\frac{\omega^{2}xr A^{+}}{6c^{2}}\left[\sin\left(2\varphi\pm\omega t\right)\right], \label{eq:gpxr} \\
g'_{x\varphi} & = &-\frac{\omega^{2}xr^{2} A^{+}}{6c^{2}}\left[\cos\left(2\varphi\pm\omega t\right)\right], \label{eq:gpxph} \\
g'_{xx} & = & 1+\frac{\omega^{2}r^{2} A^{+}}{6c^{2}}\left[\sin\left(2\varphi\pm\omega t\right)\right]. \label{eq:gpxx} 
\end{eqnarray}

In particular, in the plane $x=0$, we obtain

\begin{align}
ds^{2}&=-c^{2}\left[1-\frac{\omega^{2}r^{2}A^{+}}{2c^{2}}\sin\left(2\varphi\pm\omega t \right) \right] dt^{2}- \frac{2\omega^{2}r^{2} A^{+}}{3c^{2}}\left[\sin\left(2\varphi\pm\omega t\right)\right]dx cdt+dr^{2}+r^{2}d\varphi^{2}+\\
& +\left[1+\frac{\omega^{2}r^{2} A^{+}}{6c^{2}}\sin\left(2\varphi\pm\omega t\right)\right]dx^{2} \label{eq:metricapiano}
\end{align}

By comparing the line element (\ref{eq:weakfieldmetric11}) and metric tensor in cylindrical coordinates (\ref{eq:gp00})-(\ref{eq:gpxx}) it is possible to obtain the following expressions for the gravitoelectric potential
\beq
\Phi = \frac{\omega^{2}r^{2}A^{+}}{4} \sin\left(2\varphi\pm \omega t \right) \label{eq:phicyl}
\eeq
and for the gravitomagnetic potential

\begin{eqnarray}
A_{x} & = & \frac{1}{6} \omega^{2}r^{2}A^{+} \sin\left(2\varphi\pm \omega t \right) \label{eq:Axcyl} \\
A_{r} & = & -\frac{1}{6} \omega^{2}r x A^{+} \sin\left(2\varphi\pm \omega t \right) \label{eq:Arcyl} \\
A_{\varphi} & = & - \frac{1}{6} \omega^{2}r x A^{+} \cos\left(2\varphi\pm \omega t \right) \label{eq:Aphicyl} \\
\end{eqnarray}

Accordingly, we get the following expressions (valid up to linear order in the displacements from the origin \cite{Ruggiero:2022gzl}) for the gravitoelectric field
\beq
\mathbf E= -\bm{\nabla} \Phi = \frac 1 2 \omega^{2}r A^{+} \left[ \sin\left(2\varphi\pm \omega t \right) \mb u_{r}+ \cos\left(2\varphi\pm \omega t \right) \mb u_{\varphi}  \right] \label{eq:Ecyl}
\eeq
and gravitomagnetic field
\beq
\mb B= -\bm{\nabla} \wedge \mb A = \frac 1 2 \omega^{2}r A^{+} \left[ \cos\left(2\varphi\pm \omega t \right) \bm u_{r}- \sin\left(2\varphi\pm \omega t \right) \bm u_{\varphi}  \right]
\eeq
It is manifest that the two fields have the same magnitude $|\mb E|=|\mb B|=\frac 1 2 \omega^{2}r A^{+}$ and that they are orthogonal to each other $\mb E \cdot \mb B=0$, and orthogonal to the propagation direction. These properties were already known on the basis of the corresponding Cartesian expressions \cite{Ruggiero:2021qnu}.

 %------------------------Section-------------------------
\section{Discussion and Conclusions} \label{sec:disconc}
%------------------------Section-------------------------

We described the field of a plane gravitational wave, in the proper detector frame, using Fermi coordinates; in particular, we exploited the analogy that enables to describe this field in terms of  gravitoelectric $\Phi$ and gravitomagnetic $A_{i}$ potentials, and a perturbation of the spatial metric $\Psi_{ij}$. In this frame,  all quantities involved are relative to the corresponding ones measured at the origin of the frame, i.e. along the reference world-line, and  this fact simply points out the tidal effects produced by the passage of GWs.  The proper detector frame is a natural tool to describe the interaction of GWs with an experimental apparatus; moreover, the gravitoelectromagnetic analogy is useful to  simplify this description and to have further insights in terms of physically measurable quantities. 

In particular, we evaluated the impact of the passage of GWs on time measurements. More specifically, we considered the modification of the  paces of the clocks in the frame with respect to the reference clock at the origin: in particular, the proper time of the latter is given by construction by the time coordinate $T$.

Eq. (\ref{eq:dTfisso2}) relates the proper time interval $\Delta \tau$ measured by clock at the spatial location $X,Y,Z$ with the corresponding time interval $\Delta T$ measured by the reference clock. We see that this effect is null along the propagation direction of the wave, where $Y=Z=0$: this is a consequence of the transversality of GWs.
As for an estimate of the effect, we get
\beq
\frac{\Delta \tau-\Delta T}{\Delta T} \simeq \frac{T_{c}}{\Delta T}\frac{L}{\lambda} A^{+}. \label{eq:stima1}
\eeq
We remember that $L$ is a scale length of the reference frame and $\lambda$ is the wave length; we introduced $T_{c} = \frac{L}{c}$ which is a time scale for the propagation of light signal within the reference frame. Since $L \ll \lambda$ and on Earth it is expected that $A^{+}\simeq 10^{-20}$, we see that in any case the relative variation is very small. 

We also evaluated the desynchronization induced by the GWs, given by Eq. (\ref{eq:desynch2});  also this effect is null along the propagation direction. In order to evaluate the order of magnitude, we may write
\beq
\Delta T_{synch} \simeq \left(\frac{L}{\lambda}\right)^{2} T_{c} A^{+}. \label{eq:stima2}
\eeq
To fix the ideas, let $L \simeq 10^{6} \ \mathrm m$, so we consider two  distant places on  Earth, $\frac L \lambda \simeq 10^{-1}$, so that  the frequency of the waves is about $1 \ \mathrm{Hz}$ or smaller; if the clock accuracy is of $1 \ \mathrm{ns}$, we get $A^{+} \gtrsim 10^{-6}$. So, the effect can hardly be detected with GWs whose amplitude is much smaller.

The other situation that we considered  is referred to a clock rotating with rotation rate $\Omega$ at distance $R$ from the origin;  apart from a purely kinematic term,  the order of magnitude of the gravitational correction (\ref{eq:circ1}) is
\beq
\Delta \tau_{GW} \simeq \frac{R}{\lambda}T_{c} A^{+},\label{eq:stima3}
\eeq
if $\omega \gg \Omega$ and where, in this case $T_{c} = \frac{R}{c}$ is in the order of the rotation period.  To fix the ideas, let us consider the case of GPS satellites moving around the Earth. In this case our reference frame can be thought of as the Earth Centred Inertial (ECI) frame (and we neglect the effects due to the gravitational field of the Earth). Since $R \simeq 2 \times 10^{7}\ \mathrm m$, the long wavelength approximation tells us that we are considering waves with frequency in the order of $1\ \mathrm{Hz}$ or smaller.  We get $A^{+} \gtrsim 10^{-7}$ (using a clock accuracy of $1 \ \mathrm{ns}$) which is much larger than what we expect from GWs sources. As a comment, we add that when $2\Omega=-\mp \omega$, this effect is absent for circularly polarized GWs:  as a matter of fact, circular polarization of GWs is an important tool to test gravity and the origin of stochastic background \cite{circular}.

{In summary, we can say that even if the above described effects are in principle present, it is hard to imagine that they could  lead to an  experimental detection in the near future, taking into account the sensitivities of current devices;  more realistically, it is expected that gravitomagnetic effects could be relevant to develop new detection techniques based on the interaction of GWs with spinning particles or to characterize the binary system dynamics, as discussed in the introduction.}

In addition, we obtained the expression of the metric in the proper detector frame in  cylindrical coordinates adapted to the propagation direction of the GWs: as far as we know, this is expression was not obtained previously. In particular, we obtained the corresponding expressions of the gravitoelectric and gravitomagnetic fields, which have the form of vectors orthogonal to each other and rotating with the wave frequency $\omega$ in the plane orthogonal to the propagation direction. Eventually, it is interesting to point out that using the expression of the metric in cylindrical coordinates  given by Eqs. (\ref{eq:gp00})-(\ref{eq:gpxx}), a change to a reference frame rotating around the propagation direction with rotation rate $\Omega$ shows that in this frame the frequency of the wave is $2\Omega\pm \omega$: this is a manifestation of the coupling between the helicity of the gravitational wave and the rotation of the frame \cite{Ramos:2006sb,Mashhoon:2022fua}.  In particular, this phenomenon is already taken into account and measured for electromagnetic waves \cite{mashhoonhauck}, and is related to the so-called phase wrap-up in the GPS system \cite{Ashby:2003vja}.

\section*{Data availability}

Data sharing not applicable to this article as no datasets were generated or analysed during the current study.

\begin{acknowledgments}
The author acknowledges the contribution of  the local research  project \textit{Modelli gravitazionali per lo studio dell'universo (2022)} - Dipartimento di Matematica ``G.Peano'', Universit\`a degli Studi di Torino; this  work is done within the activity of the Gruppo Nazionale per la Fisica Matematica (GNFM). \end{acknowledgments}

%\bibliography{GEM_GW}


\end{document}